\documentclass{iopart}
\usepackage{epsfig}
\usepackage{latexsym}

\begin{document}

\newcommand{\tbox}[1]{\mbox{\tiny #1}}
\newcommand{\half}{\mbox{\small $\frac{1}{2}$}}
\newcommand{\mbf}[1]{{\mathbf #1}}

\newcommand{\cn}[1]{\begin{center} #1 \end{center}} 
\newcommand{\hide}[1]{#1} 
\newcommand{\mboxs}[1]{\mbox{\small #1}} 
\newcommand{\mpg}[2]{\begin{minipage}[t]{#1cm}{#2}\end{minipage}}	
\newcommand{\mpb}[2]{\begin{minipage}[b]{#1cm}{#2}\end{minipage}}
\newcommand{\mpc}[2]{\begin{minipage}[c]{#1cm}{#2}\end{minipage}}


\title{Overview: Brownian Motion and Dephasing due to Dynamical Disorder} 

\author{Doron Cohen}

\address{Department of Physics, Harvard University, Cambridge, MA 02138, USA}

\begin{abstract}
The motion of a particle under the influence 
of a dynamical disorder is described by the DLD model. 
One motivation is to understand the motion of an electron 
inside a metal; Another is to understand quantal Brownian motion. 
The overview is based on a research report for 1996-1998.
\end{abstract}

A generic toy model for quantal Brownian motion that takes into 
account the disordered nature of an environment (See Fig.1) has been 
defined and explored [P1-P5].   
The treatment of diffusion localization and dissipation (DLD) has been  
unified, the propagator of the reduced probability-density-matrix 
has been calculated, and its non-classical structure has been explained. \\

One motivation for studying this effective `DLD model', which constitutes 
a non-trivial generalization of Zwanzig-Caldeira-Leggett (ZCL) model,   
is the wish to understand the motion of an electron inside a metal, 
taking into account both the the static disorder configuration and also   
the Coulomb interaction with the rest of the Fermi sea. \\

An important issue [P3] is the study of decoherence using Wigner's 
phase-space representation for the description of the evolving 
quantum-mechanical state. In case of the ZCL model the propagator 
of the Wigner function is just a Gaussian kernel. In case of 
the DLD model the propagator contains a singular term  
that corresponds to an unscattered component of the wavepacket.   
Consequently it is possible to distinguish between 
{\em smearing mechanism} and {\em scattering mechanism} 
for decoherence. This distinction is essential in order to 
get a proper understanding of {\em dephasing}. \\

The extension of the latter study [P4] to the low-temperatures 
regime has been done in collaboration with {\em Y. Imry}.  
The limitations of the semiclassical strategy have been clarified. 
The work was motivated by a controversy regarding the 
effect of `zero point fluctuations'~[1],
following experimental observation 
by {\em Mohanty}, {\em Jariwala} and {\em Webb}~[2]. \\

The study of dephasing has been extended [P3] to various types 
of transport, which are illustrated by Fig.2. The main goal  
was to derive results for all these cases using a general formula 
for dephasing that applies to any temperature. 
The final result can be written in terms of two functions: 
the form-factor of the environment and the power spectrum 
of the motion under consideration. The introduction 
of ad-hoc or ambiguous cutoffs into the calculations,  
as in the works of Chakravarty and Schmid~[3]
and followers, is not required.

\newpage

\mbox{\epsfysize=4in \epsffile{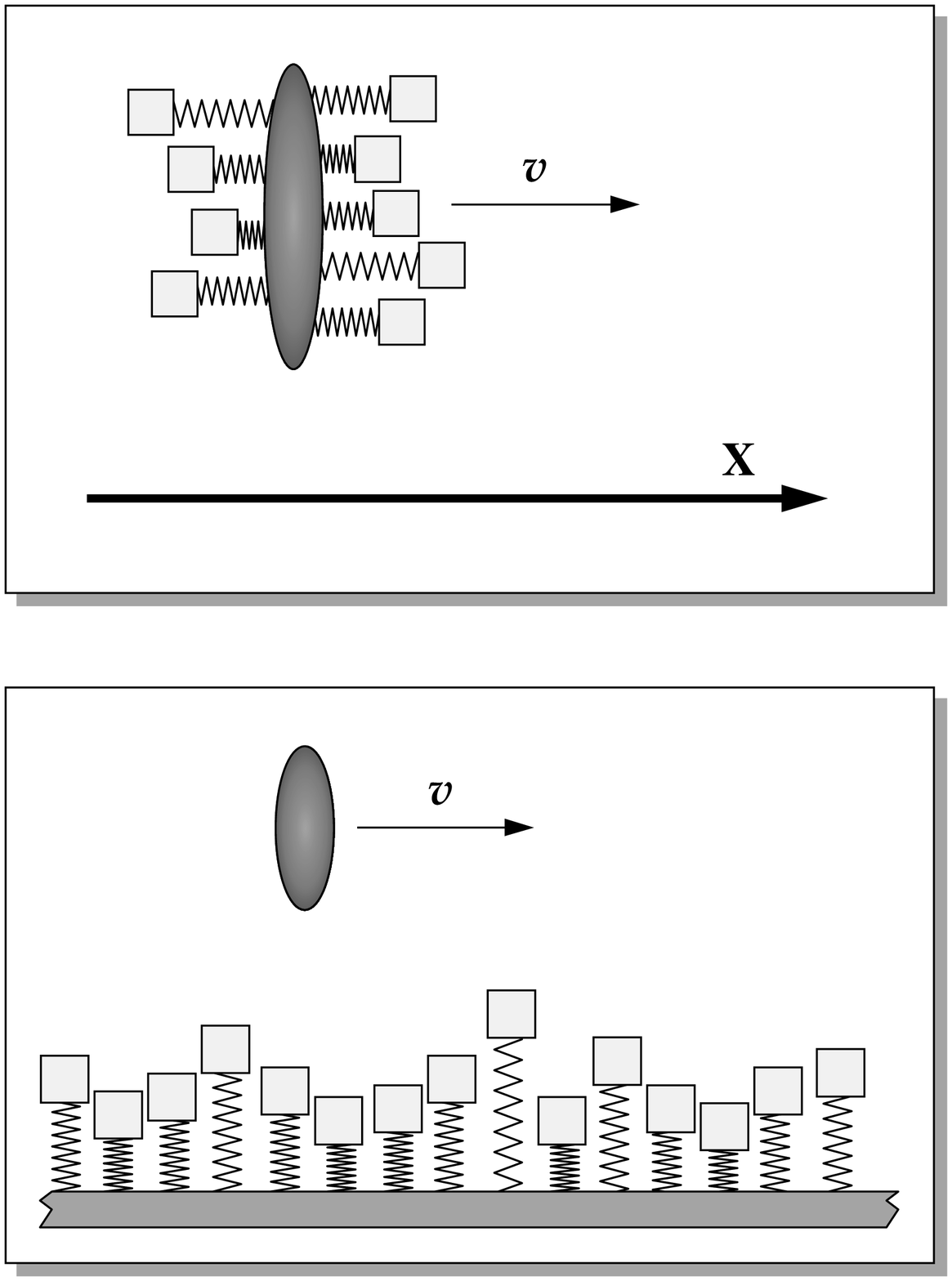}} 
\mpb{3}{
\mbox{\hspace*{0.4cm} 
\epsfxsize=2.0in \epsfysize=1.5in \epsffile{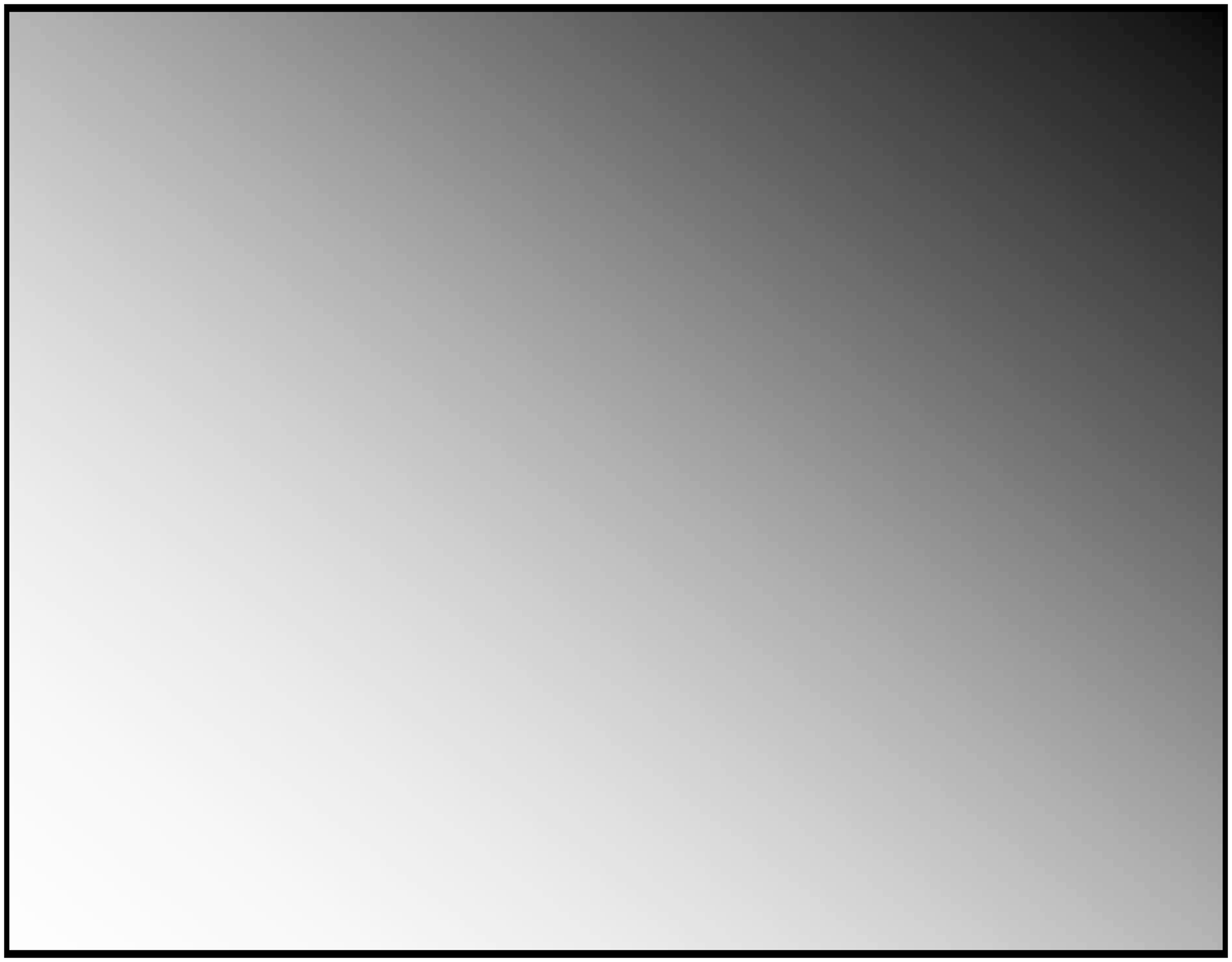}} \\
\vspace*{-0.5cm}
\mbox{\epsfysize=2.15in \epsffile{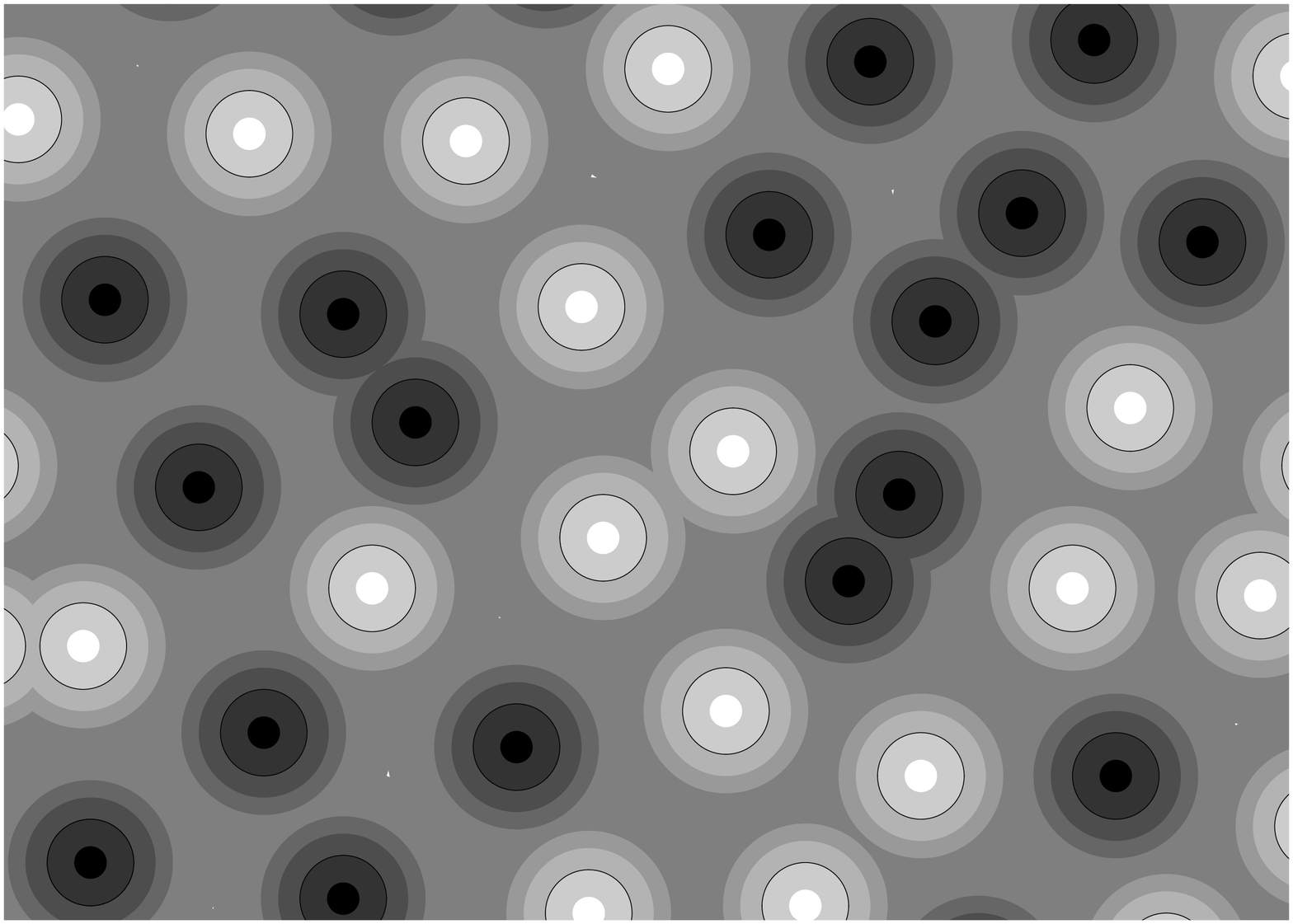}} 
} 

{\footnotesize {\bf Fig.1.} Illustration of the Zwanzig-Caldeira-Leggett 
model (upper drawings), versus the DLD model (lower drawings). 
DLD stands for `Diffusion Localization and Dissipation', which are the 
three main dynamical effects that are associated with the motion in 
dynamical disorder.  The instantaneous potential that is experienced 
by the particle is either linear (right upper drawing), 
or of disordered nature (right lower drawing) respectively. 
If the fluctuations are uncorrelated in time, then the 
two models are classically equivalent. There is no such equivalence 
in the quantum-mechanical case!} 

\ \\ \ \\ \ \\

\begin{center}
\leavevmode 
\epsfysize=1.4in
\epsffile{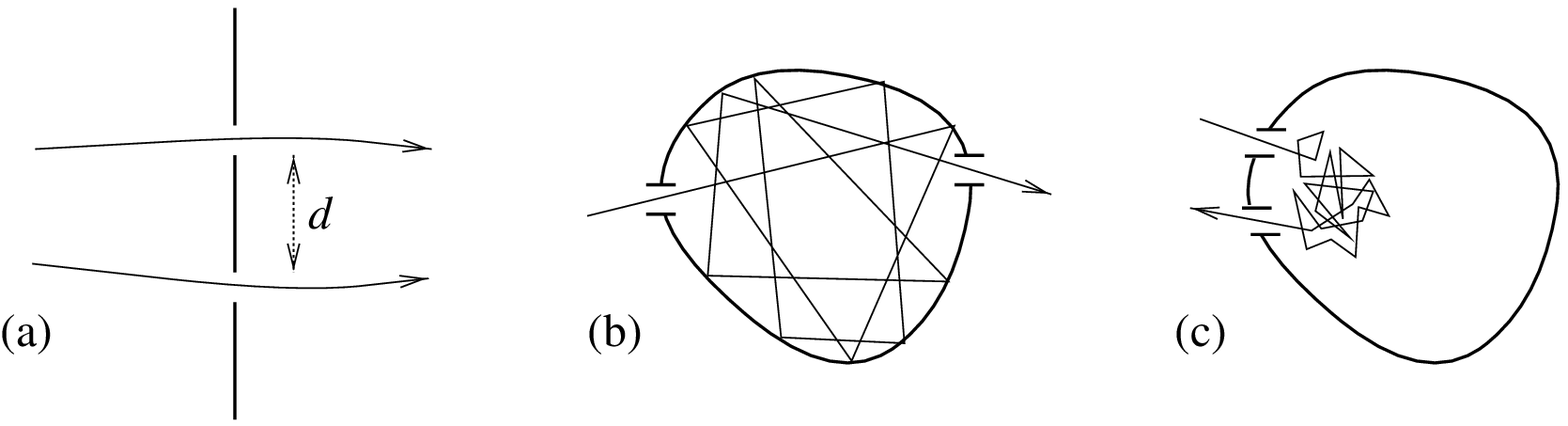}
\end{center}
{\footnotesize {\bf Fig.2.}
Various types of transport problems that have been studied: 
(a) Ballistic transport as in the two-slit experiment; 
(b) Transport via a chaotic cavity;  (c) Transport via diffusive 
cavity, as in weak localization experiments. 
The dephasing factor that suppresses the interference contribution 
can be determined once the properly-defined power-spectrum 
of the motion is calculated.} 

\newpage

\ \\ \ \\ \ \\
{\bf References:}

\begin{description}

\item[{\rm [1]}]
D.S. Golubev and A.D.Zaikin, 
{\em Phys. Rev. Lett.} {\bf 81}, 1074 (1998).

\item[{\rm [2]}]
P. Mohanty, E.M. Jariwala and R.A. Webb, 
{\em Phys. Rev. Lett.} {\bf 77}, 3366 (1997).

\item[{\rm [3]}]
S. Chakravarty and A. Schmid, {\em Phys. Rep.} {\bf 140}, 193 (1986). 

\end{description}

\ \\ \ \\
{\bf Publications:}

\begin{description}

\item[{\rm [P1]}]
D. Cohen, "Unified Model for the Study of Diffusion 
Localization and Dissipation",
{\it Phys. Rev. E {\bf 55}}, 1422-1441 (1997).
 
\item[{\rm [P2]}]
D. Cohen, "Quantum Dissipation versus Classical Dissipation 
for Generalized Brownian Motion", 
{\it Phys. Rev. Lett. {\bf 78}}, 2878 (1997).

\item[{\rm [P3]}]    
D. Cohen, "Quantal Brownian Motion - Dephasing and Dissipation", 
{\it J. Phys. A {\bf 31}}, 8199-8220 (1998). 

\item[{\rm [P4]}]    
D. Cohen and Y. Imry, "Dephasing at Low Temperatures", 
{\it Phys. Rev. B {\bf 59}}, 11143-11146 (1999).

\item[{\rm [P5]}]        
D. Cohen, "Chaos, Dissipation and Quantal Brownian Motion", 
Proceedings of the International School of Physics 
`Enrico Fermi' Course CXLIII ``New Directions in Quantum Chaos'', 
Edited by G. Casati, I. Guarneri and U. Smilansky, 
IOS Press, Amsterdam (2000). 

\end{description}

\ \\
\ \\ 
\rule{10cm}{.01in}

\noindent
{\footnotesize Preprints and publications are available via \\
{\em http://monsoon.harvard.edu/$\sim$doron}}

\end{document}